# Enhancement of piezoelectric response in V doped LiNbO$_3$ films deposited by RF magnetron sputtering

Xiaomei Zeng[1,2], Ting Lv[1], Xiangyu Zhang[2], Zhong Zeng[2], Bing Yang[2], Alexander Pogrebnjak[4,5], Vasiliy O. Pelenovich[1,2,3*], Sheng Liu[1,2,3]

[1] The Institute of Technological Sciences, Wuhan University, Wuhan, 430072, China
[2] School of Power and Mechanical Engineering, Wuhan University, Wuhan, 430072, China
[3] Hubei Key Laboratory of Electronic Manufacturing and Packaging Integration, Wuhan University, Wuhan, 430072, China
[4] Faculty of Electronics and Information Technology, Sumy State University, 40007 Sumy, Ukraine
[5] Institute of Materials Science, Slovak University of Technology in Bratislava, 917 24 Trnava, Slovak Republic

Corresponding author: Vasiliy O. Pelenovich, v.pelenovich@whu.edu.cn

## Abstract

LiNbO$_3$ films doped with vanadium (V) were deposited using RF magnetron sputtering technique. To realize doping with a wider range of V concentration, a 30 mm V metal inlaid target asymmetrically embedded in the 150 mm lithium niobate target was used. The V concentration in the deposited films was a decreasing function of the distance from the V target. The V/Nb ratio in the film decreased from 0.155 to 0.024. Surface and inner morphology and structure, phase and element composition, microstructure, and ferroelectric properties of the undoped and V doped LiNbO$_3$ films were studied. The measured maximal $d_{33}$ constant of the LiNbVO film with V/Nb ratio of 0.07 was about three times higher than that of the undoped LiNbO$_3$ film, 13.5 pC/N and 4.76 pC/N, respectively. The optimal composition in the deposition geometry used was within the V/Nb ratio range of 0.05 to 0.13. Undoped and V doped LiNbO$_3$ thin films were used as bulk acoustic wave ultrasonic transducers deposited on stainless steel plates to generate longitudinal waves and compare their ultrasonic performance.

# 1. Introduction

The excellent piezoelectric performance of lithium niobate (LiNbO$_3$) with high Curie temperature of ~1483 K and coupling coefficient $k_t$ of 0.49 [1,2,3,4] determines its widespread use in piezoelectric devices such as micro-mechanical ultrasonic transducers [5,6,7,8,9], shear wave resonators [10], surface acoustic wave sensors [11,12], and precision positioning drivers [13].

In order to simplify structure of the ultrasonic transducer, improve its high temperature durability, and avoid problems of adhesive between transducer and sample, deposition of the piezoelectric transducer in form of thin films is preferred. In addition, thin film deposition technologically is simpler than high quality single crystal growth, thin films have stronger adhesion to substrate, and miniaturization of electronic devices is possible. To prepare LiNbO$_3$ thin films, various techniques can be used, such as RF magnetron sputtering [14,15], pulsed laser deposition [16,17], sol-gel method [18,19], thermal assisted spin coating [20], chemical bath deposition [21], polymeric precursor method [22], photolithographic technique [23], and ion beam sputtering [24].

In recent years, doping specific metal elements into AlN piezoelectric materials has been proven to be an effective method for improving piezoelectric properties. Akiyama et al. showed that Sc doping of AlN (Sc$_{0.43}$Al$_{0.57}$N alloy) exhibited a large piezoelectric coefficient $d_{33}$ of 27.6 pC/N, which was 500% larger than that of undoped AlN [25]. Similarly, there were made multiple attempts to improve the piezoelectric performance of LiNbO$_3$ in bulk form (crystals and powders) by doping with Mg, B, and Zn. Compared with undoped LiNbO$_3$, LiNbO$_3$ single crystals [26,27,28] and powders [29] doped with Mg up to 10 mol.% demonstrated not significant improvement or deterioration in piezoelectric properties. While LiNbO$_3$ single crystals doped with 5.6 mol.% Zn and 0.44 wt.% Gd after a certain heat treatment showed an increase in piezoelectric $d_{33}$ constant to the maximum values of 13.2 pC/N and ~12 pC/N, respectively, which are close to that of the single domain pure LiNbO$_3$ single crystals [30,31].

However, in literature, there is little research on study of ultrasonic response of doped LiNbO$_3$ thin films. Nakamura et al. deposited *c*-axis oriented pure LiNbO$_3$ and Pr doped LiNbO$_3$ films using RF magnetron sputtering. The electromechanical coupling coefficient $k_t^2$ demonstrated increase from 0.8% of the pure film up to 2.2% of the Pr doped film [32], which is closer to the theoretical value of 4% in the *c*-axis (Z-cut) oriented LiNbO$_3$ crystal. Recently, Kobayashi et. al prepared 100 μm thick LiNbO$_3$/TiO$_2$+SrCO$_3$ composite films using sol-gel technique and observed high temperature ultrasonic response up to 1050 °C [19]. The other composites, LiNbO$_3$/TiO$_2$ and LiNbO$_3$/Al$_2$O$_3$, demonstrated lower temperature durability up to 700 °C [18].

Though the LiNbO$_3$ is high temperature ferro- and piezoelectric material, its $d_{33}$ strain constant of ~6 pC/N is very low in comparison with those of other piezoelectric materials [1,2]. In our previous study, we found that the $d_{33}$ constant of LiNbO$_3$ in film form depends on the thickness, but cannot exceed 6 pC/N. In this study, we characterize LiNbO$_3$ thin films doped with vanadium (V) deposited using magnetron sputtering technique. The vanadium doping greatly improves the piezoelectric constant of LiNbO$_3$ piezoelectric films. Effects of V concentration on piezoelectric performance are studied. We also apply V doped LiNbO$_3$ thin films for bulk acoustic wave ultrasonic transducers to generate longitudinal waves.

## 2. Results and discussion

In order to reach a wider range of V concentration in the LiNbO$_3$ films an asymmetrically embedded V inlaid target was used, as shown in Fig. 1a and 1b. To prepare the inlaid target, a 150 mm target containing LiNbO$_3$ and LiNb$_3$O$_8$ phases was drilled at a radial distance of 50 mm, with a hole diameter of 30 mm. The 30 mm V inlaid target was installed in this hole. The rectangular Si and SS substrates were placed relative to the V target, as shown in Fig. 1a. In this geometry the V concentration in RF plasma is a decreasing function of the distance from the V target, as shown in Fig. 1c for V/Nb atomic ratio in the deposited LiNbO$_3$ film transducers measured by energy dispersive spectroscopy (EDS) technique. As reference samples, pure undoped LiNbO$_3$ film transducers were deposited using the same 150 mm target with an 30 mm inlaid target of the same material (mixture of LiNbO$_3$ and LiNb$_3$O$_8$ phases).

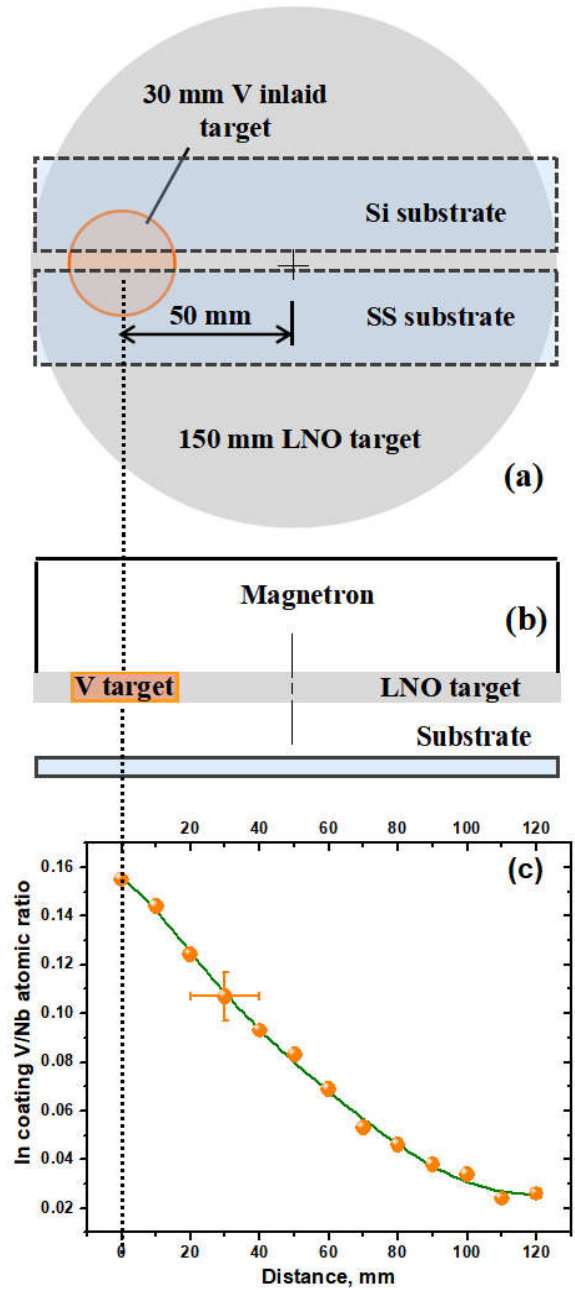

**Fig. 1**. Schematic diagrams of deposition geometry, plane view (a) and cross-section (b). The dependence of V/Nb atomic ratio in deposited films on distance from the V inlaid target (c). The distance scale in all figures corresponds to the *x*-axis in figure (c).

In Figs. 2 surface and cross-section morphology of the undoped LiNbO$_3$ film (Figs. 2a and 2b) and V doped LiNbO$_3$ film with V/Nb atomic ratio of 0.07, i.e. at a distance from the inlaid target of 60 mm (Figs. 2d and 2e) are shown. In the cross-section image of the undoped film a well developed columnar structure is observed. Therefore, each particle in Fig. 2a represents the top of the column. The V doped LiNbO$_3$ film has finer and less oriented columnar structure with featureless, probably amorphous, bottom layer of ~5 μm in thickness. The total thickness of the deposited transducers in the deposition center and at the edge is ~24 μm and ~15 μm, respectively. In Figs. 2c and 2f AFM images of undoped LiNbO$_3$ and V doped LiNbO$_3$ films deposited on SS substrate, respectively, are shown. The roughness of the undoped and V doped films measured on an area of 10 × 10 μm$^2$ is 31±3 nm and 99±8 nm, respectively.

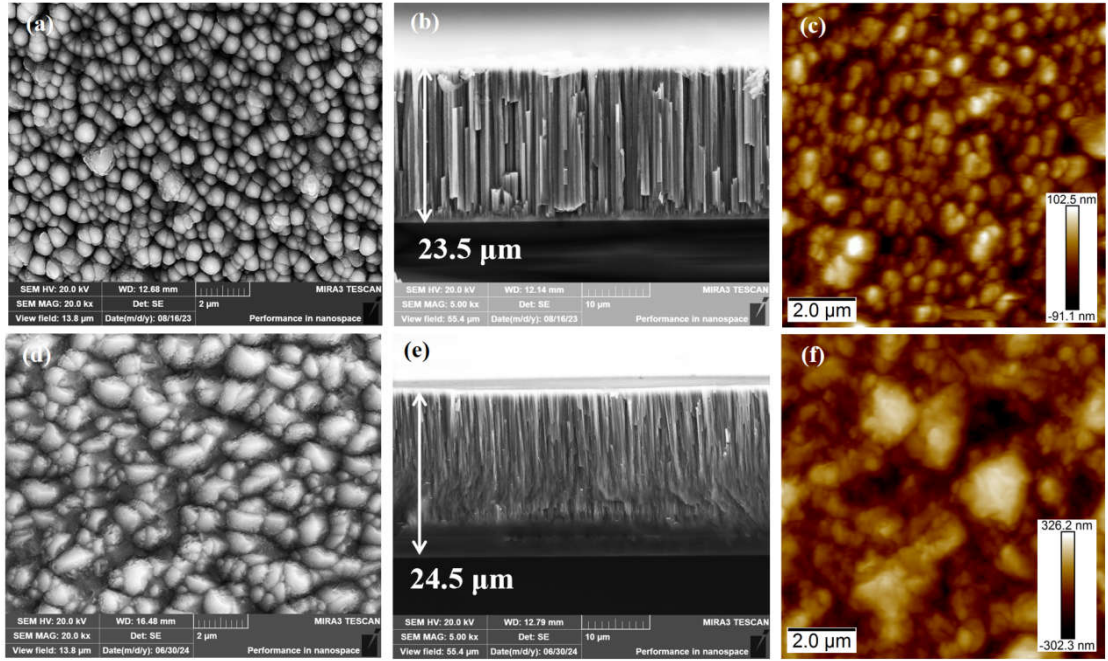

**Fig. 2**. SEM images of surface and cross-section of undoped LiNbO$_3$ film (a, b) and V doped LiNbO$_3$ film with V/Nb atomic ratio of 0.07 (d, e). AFM images of undoped LiNbO$_3$ film (c) and V doped LiNbO$_3$ film (f)

In Fig. 3a XRD patterns of undoped and V doped LiNbO$_3$ films are shown. Undoped film contains two phases, the LiNbO$_3$ and Li-deficient piezoelectrically inactive LiNb$_3$O$_8$ phase with (012) and (116) preferred reflections, respectively. V doped films demonstrate more polycrystalline LiNbO$_3$ phase with dominant (012) reflection and traces of LiNb$_3$O$_8$ phase.

We can use (012) reflection of the LiNbO$_3$ phase to calculate the mean grain size $D$ by the Scherrer equation:

$$D = \frac{K\lambda}{\beta \cos\theta}$$

where $K \sim 0.9$ is a shape factor, $\lambda = 0.154$ nm is the X-ray wavelength, $\beta$ is the full width in radian of a certain reflection at its half maximum of intensity (FWHM), and $\theta$ is the Bragg angle of the (012) reflection equal to 23.3/2 deg. The estimated mean grain sizes of undoped and V doped $LiNbO_3$ films are 54±2 nm and 33±3 nm, respectively. The smaller grain size of the V doped $LiNbO_3$ film is in agreement with the observation of finer columnar structure in the cross-section SEM image.

In Fig. 3b unpolarized Raman spectra of undoped $LiNbO_3$ film and V doped $LiNbO_3$ film with V/Nb atomic ratio of 0.07 are shown. The undoped $LiNbO_3$ film exhibits broad peaks, which were identified with $LiNbO_3$ and $LiNb_3O_8$ phases [33] (blue line). The V doped $LiNbO_3$ film demonstrates the main peaks of the $LiNbO_3$ phase and traces of the $LiNb_3O_8$ phase as a shoulder at 688 cm$^{-1}$ of the main $LiNbO_3$ peak at 620 cm$^{-1}$ (orange line). All obtained results are in agreement with the XRD data. Figs. 3c and 3d show the Raman spectra of undoped and V doped film measured on cross-section at different depths, respectively. Fig. 3e shows the corresponding cross-section and positions of the measurement points (V doped film). Both undoped and V doped films demonstrate similar phase composition independent of the depth. Therefore, the deposited V doped film demonstrates phase homogeneity independent of the film depth.

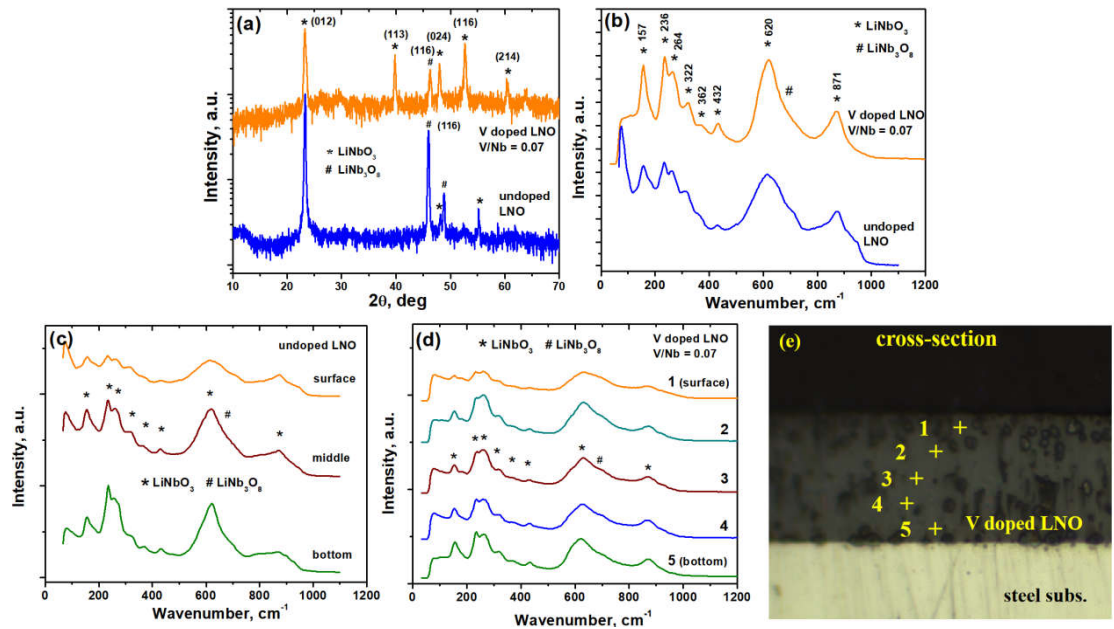

**Fig. 3**. XRD patterns (a) and unpolarized Raman spectra (b) of undoped $LiNbO_3$ film and V doped $LiNbO_3$ film with V/Nb atomic ratio of 0.07. Cross-sectional Raman spectra of undoped (c) and V doped films (d) measured at different film depth and corresponding cross-section optical image and measurement points (e).

The SEM-EDS mapping analysis of the undoped LiNbO$_3$ film and V doped LiNbO$_3$ film is shown in Figs. 4a and 4b, respectively. In both surface and cross-section maps all films demonstrate homogeneous distribution of Nb, O and V elements in agreement with Raman spectroscopy results.

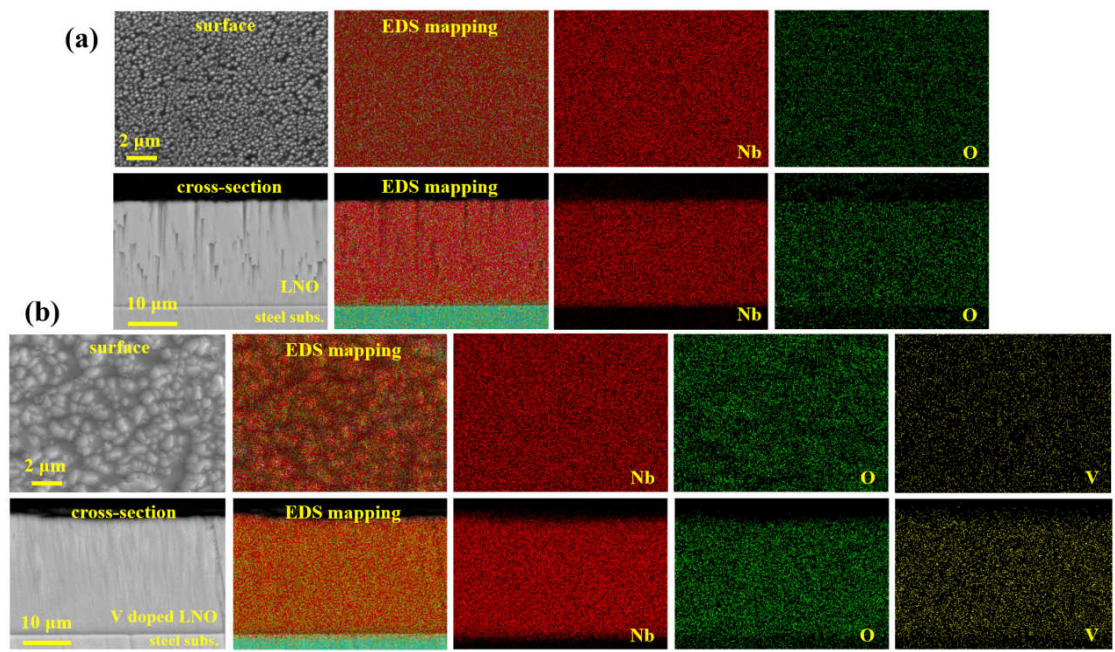

**Fig. 4**. SEM-EDS Nb, O, and V mapping of surface and cross-section of undoped LiNbO$_3$ film (a) and V doped LiNbO$_3$ film with V/Nb atomic ratio of 0.07 (b).

Fig. 5a shows a bright-field cross-sectional TEM image of V doped LiNbO$_3$ film. The film composes of columns with width in the range of 200-400 nm. In Figs. 5b and 5c HRTEM images of two columns are shown. The columns are single crystals with arbitrary orientations, which can be seen in Fast Fourier Transform (FFT) images of the selected area in the insets, in agreement with the XRD results. The chemical composition of the film revealed by TEM-EDX micro-analysis is: O of 68.1 at.%, Nb of 23.6 at.%, V of 1.25 at.%. The EDX spectrum is shown in Fig. 5c. There is also C of ~6 at.%, probably due to presence of vacuum pump oil vapor in the chamber. The V/Nb atomic ratio is 0.053, which is slightly lower than 0.07 obtained by the SEM-EDS technique. In Fig. 5e Nb, O, and V element EDX mapping of a few columns is shown. It can be concluded that the film has homogeneous distribution of Nb, O and V elements, consistent with SEM-EDS results, but now at the sub micron scale.

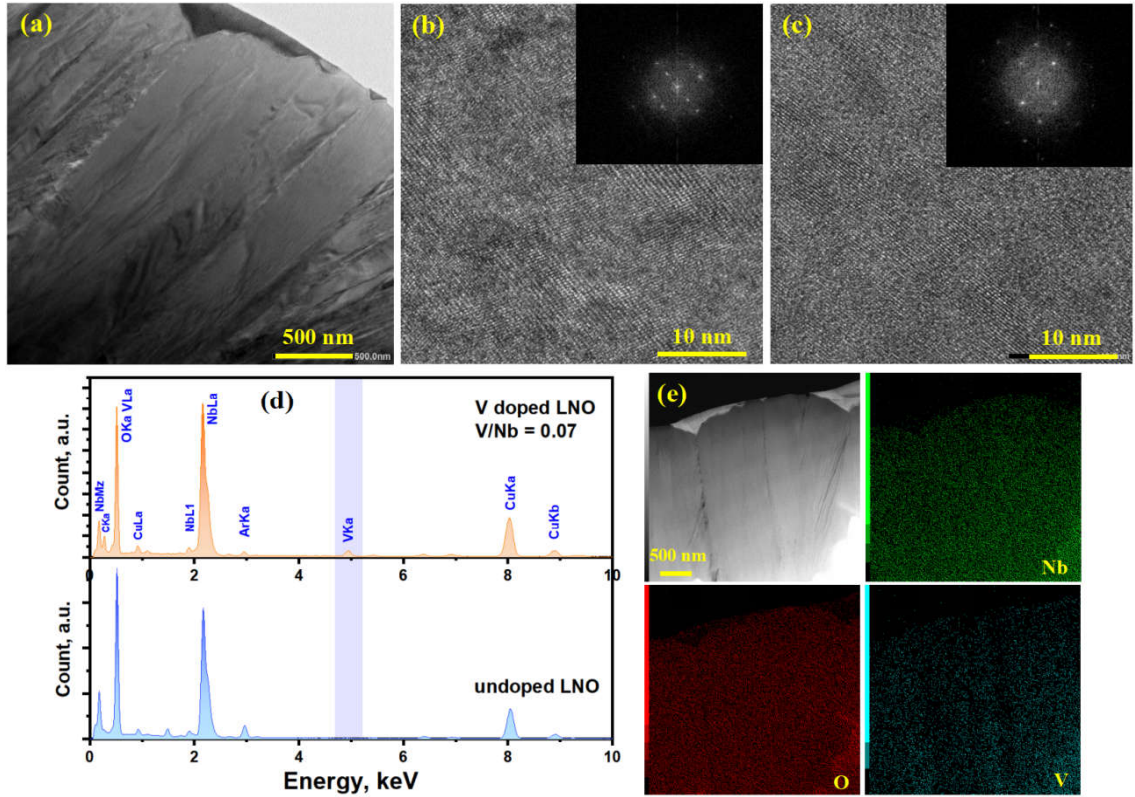

**Fig. 5.** TEM cross-section image (a) and TEM-EDX spectrum (b) of V doped LiNbO$_3$ film with V/Nb atomic ratio of 0.07. HRTEM images of two columns and (FFT) images of the selected area in the insets (c) and (d).

In Fig. 6a $d_{33}$ constants of undoped and V doped LiNbO$_3$ films as functions of distance from the inlaid target are shown. $d_{33}$ constant of the undoped LiNbO$_3$ film demonstrates significant dependence, changing from 1.68 pC/N at the deposition edges to 4.76 pC/N at the deposition center (distance from the inlaid target of 60 mm). The increase in the $d_{33}$ constant at the deposition center can be related to the higher crystal quality of the thicker films, as it was found in our previous study [35]. The V doped LiNbO$_3$ film demonstrates similar increase in the $d_{33}$ constant from 0.4 to 13.5 pC/N, when the distance from the V target increases from 0 to 60 mm. The maximal value of the $d_{33}$ constant is about three times higher than that of the undoped LiNbO$_3$ film. When the distance further increases to 120 mm, the $d_{33}$ constant decreases and tends towards the undoped LiNbO$_3$ values at the same distance, which can be explained by the fact that the concentration of V element at long distances is too low to produce any effect.

However, the distance dependence of the $d_{33}$ constant depends on the deposition geometry, which is inconvenient for practical use. Using the distance dependence of V/Nb atomic ratio (Fig. 1c), we redrew the $d_{33}$ constant as a function of V/Nb atomic ratio, as shown in Fig. 6b. The highest $d_{33}$ = 13.5 pC/N is observed at V/Nb atomic ratio of 0.07.

The obtained dependence of the $d_{33}$ constant includes two effects. First is the V/Nb atomic ratio dependence itself, representing effect of the composition on piezoelectric response. The second can be called the "geometrical" effect, which includes the mentioned thickness dependence and effect of the crystal orientation, as was found in our previous study on undoped $LiNbO_3$ film transducers [33]. In other words, although the asymmetric geometry with the inlaid target used expands the range of V concentration in the films, it does not allow to distinguish effects of doping concentration and thickness/orientation of the deposited films on ultrasonic response. Since both undoped and doped films were deposited in similar conditions, we can minimize the "geometrical" effect by calculating the ratio of $d_{33}$ constants of V doped and undoped $LiNbO_3$ films as function of V/Nb atomic ratio, as shown in Fig. 6c. The maximum now is flatter and wider and in the V/Nb atomic ratio range of 0.05 - 0.13 the $d_{33\text{LNO:V}}/d_{33\text{LNO}}$ ratio is about 3, demonstrating optimal range of the V concentration.

To independently demonstrate the increase in $d_{33}$ constant, we measured the ultrasonic pulse-echo response of V doped $LiNbO_3$ film transducers. In Fig. 6d the ultrasonic responses of undoped and V doped $LiNbO_3$ film transducers deposited at a distance of 60 mm from the inlaid target are shown. Compared to the amplitude of the undoped transducer, doping with V at a V/Nb atomic ratio of 0.07 results in an increase of over 9 times in the amplitude of the longitudinal wave echo $L_2$. In Fig. 6e distance dependence of the $L_2$ echo is shown. Similar to $d_{33}$ constant, the longitudinal wave amplitudes show their maximum values at the deposition center. In a wide distance range, the amplitude generated by V-doped $LiNbO_3$ transducer is ~9 times larger than that of the undoped $LiNbO_3$ transducer.

Next, we have measured in-situ high-temperature pulse-echo response of the V doped $LiNbO_3$ transducer deposited on the SS substrate in the temperature range of 30-600 °C, as shown in Fig. 6f. As the temperature increases, the wave echoes ($L_2$-$L_6$) shift towards higher time-of-flight. The shift is explained by increase of the thickness of the SS substrate due to the thermal expansion and by decrease of the speed of sound [36]. As the temperature increases, the wave echo amplitudes decreases, this effect can be explained by a decrease in resistivity of the transducer due to its semiconducting properties, as found in our previous study [37] on undoped LiNbO$_3$ films.

To study ferroelectric properties of the transducers we have measured ferroelectric Curie temperature using DTA technique. The undoped and V doped films demonstrate an endothermic peak at 1169 °C and 1161 °C, respectively, as shown in Fig. 6g, which can be related to the ferroelectric-paraelectric transition of the LiNbO$_3$ lattice. It can be concluded that V doping does not change significantly the ferroelectric properties of the LiNbO$_3$.

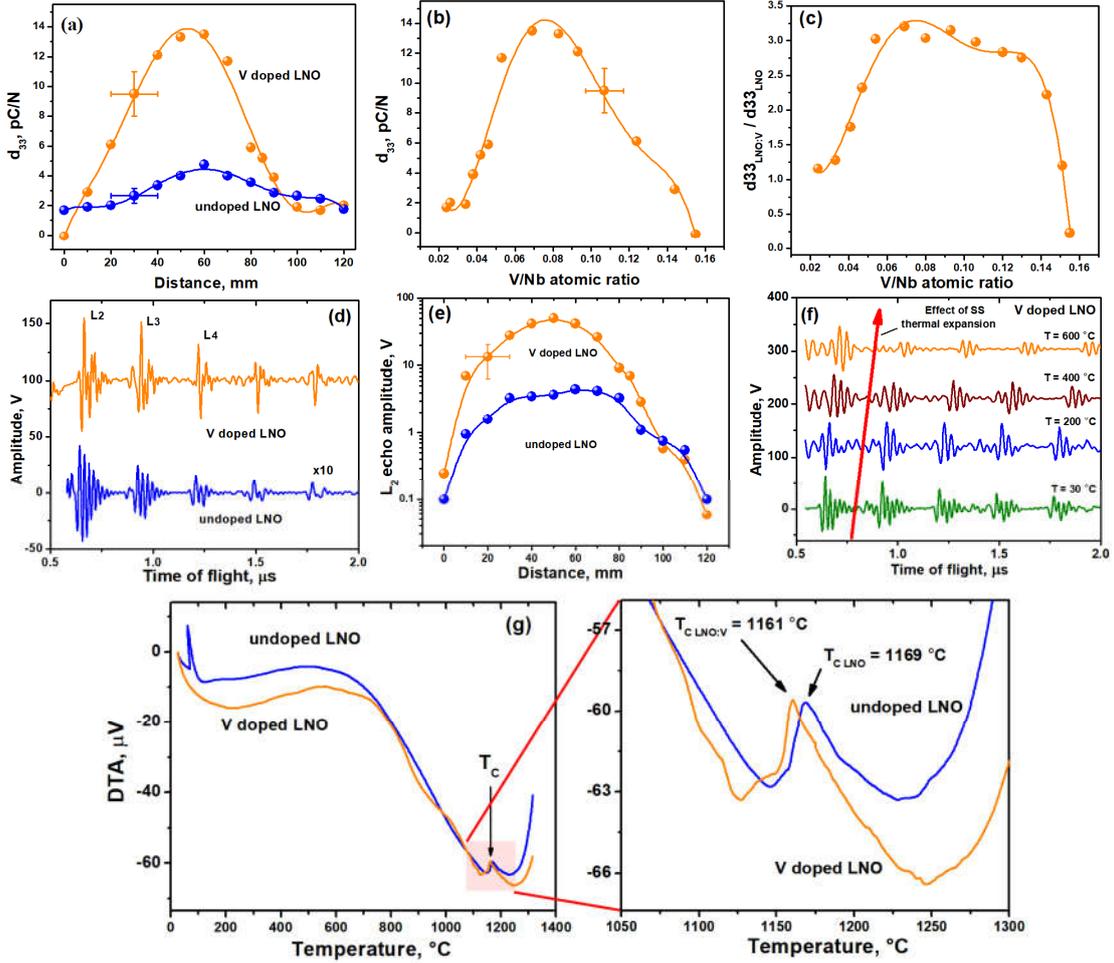

**Fig. 6**. $d_{33}$ constants of undoped and V doped (V/Nb = 0.07) LiNbO$_3$ films as functions of distance from the V inlaid target (a) and function of V/Nb atomic ratios (b). $d_{33}$ constant ratio of undoped and V doped LiNbO$_3$ films as function of V/Nb atomic ratio (c). The pulse-echo response (longitudinal waves) of undoped and V doped LiNbO$_3$ film transducers (d). The amplitudes of L$_2$ echo of undoped and V doped LiNbO$_3$ films as functions of distance from the inlaid target (e). Wave-forms of multiple echoes of V doped LiNbO$_3$ film transducer at different temperatures within the range of 30-600°C (f). DTA curves of undoped and V doped LiNbO$_3$ films and ferroelectric-paraelectric transitions near the $T_c$ (g).

The increase in $d_{33}$ constant of the V doped LiNbO$_3$ films up to 13.5 pC/N is unlikely due to possible unipolarity of the deposited films [30,31], since the XRD pattern demonstrates their polycrystalline structure (Fig. 3a). In addition, the deposited V doped LiNbO$_3$ films are even more polycrystalline than the textured undoped LiNbO$_3$ films with (012) preferred reflection. The increase in $d_{33}$ constant of doped films probably related to the difference between V$^{5+}$ and Nb$^{5+}$ cation displacements. K. Toyoura et al. using first principle theoretical investigation showed that larger Nb$^{5+}$ cation displacement in LiNbO$_3$ in comparison with Ta$^{5+}$ cation displacement in LiTaO$_3$ results in higher Curie temperature and spontaneous polarization of LiNbO$_3$ [38]. Therefore, a possible larger V$^{5+}$ cation displacement may lead to stronger spontaneous polarization, resulting in increase in the $d_{33}$ constant.

## 3. Experimental details

Piezoelectric LiNbO$_3$ (LNO) film transducers undoped and doped with V were deposited on Si (100) and 304 stainless steel (SS) substrates using RF magnetron sputtering technique. The optimized deposition parameters of the film transducers were as follows: target-substrate distance of 50 mm, radial distance from the inlaid target in the target plane within the range of 0-120 mm, RF power of 900 W, gas pressure of 1 Pa, Ar to O$_2$ flow ratio of 6/1, deposition time of 12 h, substrate temperatures in the deposition center and opposite of the V target 340 °C and 290 °C, respectively. A bias of -8 V was applied to improve morphology and structure uniformity of the deposited films. Fig. S1 shows the data of optimizing deposition parameters (gas pressure, Ar to O$_2$ flow ratio, deposition time, and bias voltage) to obtain the strongest piezoelectric response (strain constant $d_{33}$ and pulse echo response). The schematic diagrams of the deposition system and deposition geometry are shown in Fig. 7 and Fig. 1a, respectively.

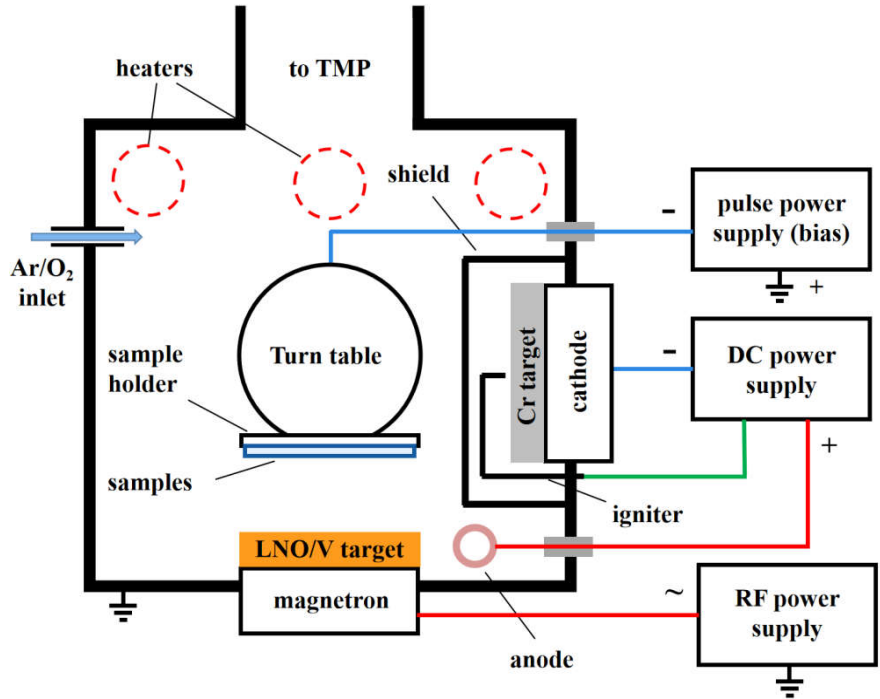

**Fig. 7**. Schematic diagram of the film deposition system.

Crystalline phase analysis was done using XRD technique (Tongda TDM-10 X-ray diffractometer with Cu K$_\alpha$ source). The surface morphology and cross-sections of the deposited films were examined by a scanning electron microscope (SEM) TESCAN MIRA3. The surface morphology and RMS roughness were measured by an atomic force microscope (AFM) Shimadzu SPM-9500 J3, operated in tapping mode with a measuring area of 10×10 μm$^2$. V/Nb atomic ratio was studied by an Aztec Energy X-Max 20 EDS analyzer attached to the SEM. Unpolarized room temperature Raman spectra were collected using TESCAN MIRA SEM with Raman attachment with an excitation wavelength of 532.08 nm. The microstructure of V doped LiNbO$_3$ film transducers was studied using a JEOL JEM-F200(HR) high-resolution transmission electron microscope (HRTEM). The HRTEM samples were prepared using the focused ion beam technique in a TESCAN GAIA3 SEM. The Curie temperatures of undoped and V doped LiNbO$_3$ films were measured by a Termo Gravimetry Differenfial Thermal Analyzer (TG/DTA) HITACHI STA7300 at a rate of 20 °C/min in the temperature range of 30 °C–1300 °C in N$_2$ atmosphere using a 70 μl Al$_2$O$_3$ crucible. The piezoelectric strain constant $d_{33}$ were measured using a SA1303A d33 Meter tester.

The ultrasonic response of the film transducers deposited on the SS substrates was measured by a DPR300 ultrasonic pulser-receiver (JSR Ultrasonics), as shown in Fig. 8. Before the measurement, 3 mm electrodes on the film were made using silver paste. The SS substrate was used as the bottom electrode. Negative spikes with an amplitude of -100 V and a length of 30 ns were used to generate ultrasonic waves using converse piezoelectric effect, as shown in Figs. 8a and 8b. The ultrasound echos reflected from the bottom of the substrate (Fig. 8c) were converted into electrical signals by the transducer direct piezoelectric effect (Fig. 8d), and then detected by the pulser-receiver with a gain of 27 dB, a low pass filter of 1 MHz, and a high pass filter of 60 MHz. The data was collected and processed by LabVeiw code (Fig. 8e).

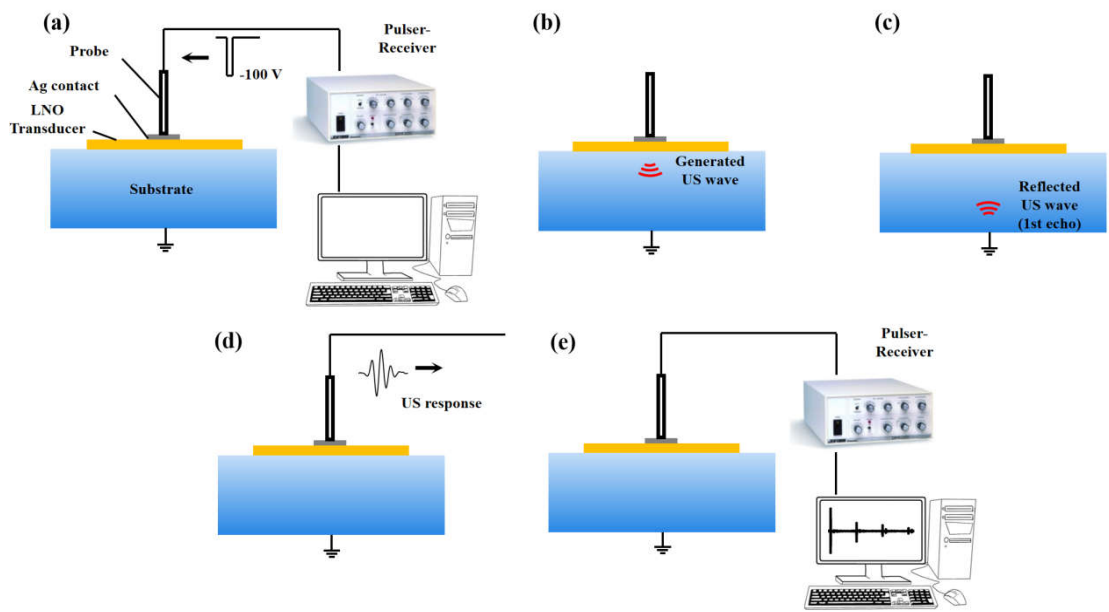

**Fig. 8**. Schematic diagram of ultrasonic response measurement. A negative electric spike is applied to the transducer (a). An ultrasonic wave is generated and propagates in the SS substrate (b). The ultrasonic wave is reflected from the substrate bottom (c). The ultrasonic wave is transformed by the transducer to the electrical signal (response) (d). Data processing by LabVeiw code (e).

## 4. Conclusions

In this study, LiNbO$_3$ film ultrasonic transducers doped with vanadium (V) were deposited using RF magnetron sputtering technique. In order to reach a wider range of V concentration in the LiNbO$_3$ films we used 30 mm V inlaid target asymmetrically embedded in the 150 mm lithium niobate target. The V/Nb atomic ratio in the deposited films was a decreasing function of the distance from the V target from 0.155 at 0 mm to 0.024 at 120 mm. The $d_{33}$ constant measurement of the doped films revealed a significant increase in its value. The highest $d_{33}$ = 13.5 pC/N is observed at V/Nb atomic ratio of 0.07, which is three times higher than 4.76 pC/N of the undoped $LiNbO_3$ film deposited at the same geometry. The optimal composition in the deposition geometry used was within the V/Nb ratio range of 0.05 to 0.13. The V doped films were used as ultrasonic transducers to generate longitudinal ultrasonic waves. The echo amplitude generated by V-doped transducers was ~9 times larger than that of the undoped $LiNbO_3$ transducers, which proved the increase in $d_{33}$ constant of V doped films. The observed increase in $d_{33}$ constant explained by a stronger spontaneous polarization due to possible larger $V^{5+}$ cation displacement as compared to that of the $Nb^{5+}$ cation.

## Acknowledgments

This study was supported by the Fundamental Research Funds for the Central Universities, grant No. 4207-413100036; the First Batch of Special Young Scientists Open Fund for Academician Mao Ming's Workstation in 2024; Open Fund of Hubei Key Laboratory of Electronic Manufacturing and Packaging Integration (Wuhan University), grant No. EMPI2024017. We thank Miss Yanli Guo from the Core Facility of Wuhan University for her assistance with SEM-Raman analysis.

# Supplement materials

In order to obtain the highest values of the $d_{33}$ constant and echo amplitude of the ultrasonic response, some deposition parameters (bias voltage, Ar/O$_2$ flow ratio, gas pressure, and deposition time) were optimized, as shwon in Fig. S1. The decrease of the $d_{33}$ constant at high bias voltage (top row in Fig. S1) is related to the degradation of the columnar structure morphology due to ion bombardment damage, as found in our previous study on LiNbO$_3$ films [33]. The better piezoelectric performance at lower gas pressure and longer deposition time can be explained by a thickness dependence found in our previous study, where, due to better structure quality, thicker transducers exhibited larger $d_{33}$ constants [35].

## Bias voltage optimization

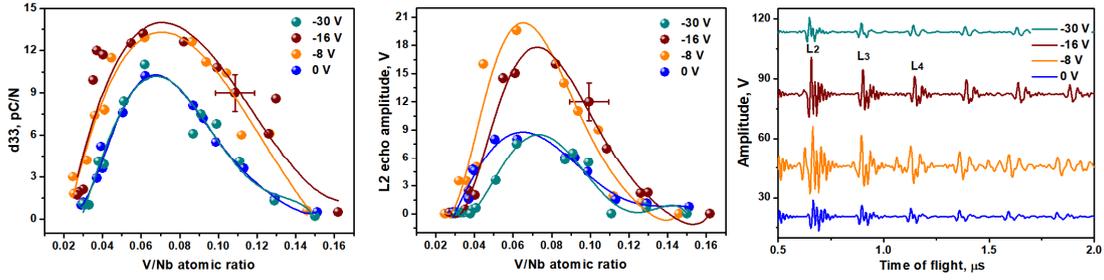

## Ar/O$_2$ gas flow ratio optimization

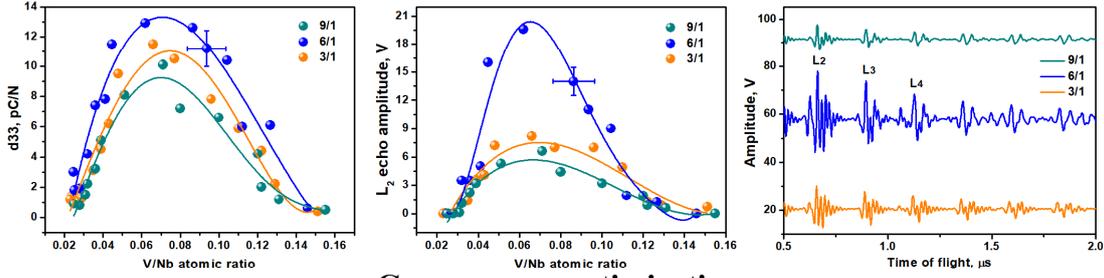

## Gas pressure optimization

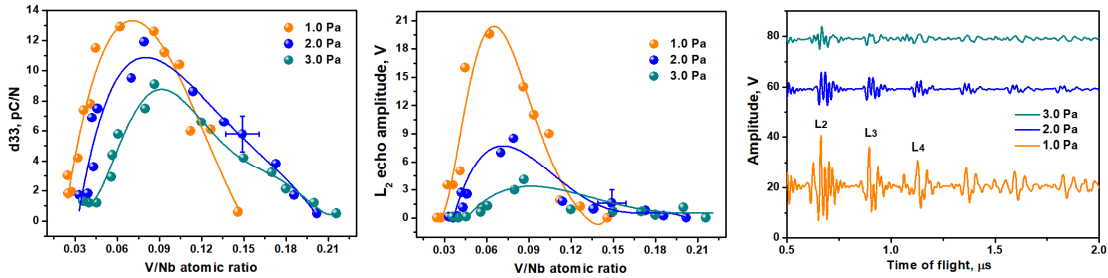

## Deposition time optimization

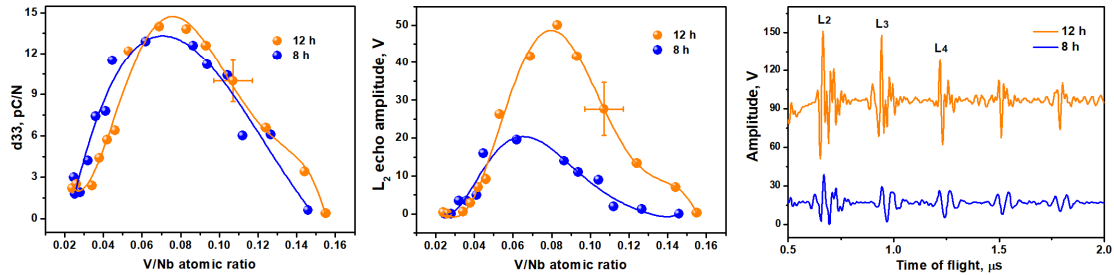

**Fig. S1**. Optimization of deposition parameters for V doped LiNbO$_3$ film transducers: bias voltage, Ar to O$_2$ gas flow ratio, gas pressure, and deposition time (from top to bottom). The *d$_{33}$* constants as functions of V/Nb atomic ratio (left column), L$_2$ echo amplitudes as functions of V/Nb atomic ratio (middle column), and pulse-echo response in the time domain (right column).